\documentclass[aps,prd,onecolumn,amssymb]{revtex4}
\usepackage{graphicx,bm,color}
\usepackage{amsmath}
\usepackage{amssymb}
\usepackage{amsfonts}
\usepackage{epsfig}
\newcommand{\be}{\begin{equation}}
\newcommand{\ee}{\end{equation}}
\newcommand{\bea}{\begin{eqnarray}}
\newcommand{\eea}{\end{eqnarray}}
\newcommand{\beaa}{\begin{eqnarray*}}
\newcommand{\eeaa}{\end{eqnarray*}}

\allowdisplaybreaks[4]

\begin{document}

\title{Quantum time in near-horizon region of a black hole}
\author{H. Hadi$^1$}\email{hamedhadi1388@gmail.com}
\author{F. Darabi$^1$}\email{f.darabi@azaruniv.ac.ir}
\author{K. Atazadeh$^1$}\email{atazadeh@azaruniv.ac.ir}
\affiliation{$^1$Department of Physics, Azarbaijan Shahid Madani University, Tabriz, 53714-161 Iran}

\begin{abstract}
The understanding of time and dynamics can be elucidated by examining the concept of entanglement in quantum theory. This particular perspective on time is referred to as the timeless approach, which posits that the universe exists in a fixed state where two separate subsystems, namely the "clock" and the "rest," are entangled. By selecting an appropriate observable for the clock, the state of the rest of the universe evolves unitarily in relation to the variable that labels the clock observable's eigenstates, which is then interpreted as time. This intriguing model, initially introduced by Page and Wootters, has also been applied to the context of curved spacetime. In this study, we explore various uncertainties pertaining to the dynamics of the rest of the universe within a curved spacetime, including ambiguities related to the clock, the system's time evolution, the flow of time, and the recording of its history. Our investigation is primarily focused on the near horizon region of a black hole, as the peculiar behavior of quantum effects in this area allows for a thorough examination of the timeless depiction proposed by Page and Wootters in describing the system's dynamics within curved spacetime. This analysis may be valuable for quantum gravity projects that align with the approach put forth by Page and Wootters. It is worth noting that the application of the Page and Wootters approach in this particular region results in a distinct clock without any ambiguity. However, the other aforementioned issues, unlike those resolved in the realm of quantum mechanics, persist in this region. 

\end{abstract}
\maketitle

\section{Introduction}
The Schr\"odinger equation considers time as a classical parameter, which sets it apart from other dynamic features like position or momentum. In the laboratory, a classical clock is used to interpret time physically. While this approach is sufficient for practical purposes, it falls short when dealing with quantum phenomena that have nontrivial features. To fully describe time in a quantum context, we need to develop a quantum description of time. Several proposals have been put forward, but there are still some controversial issues surrounding them. These proposals can be found in \cite{1,2,3,4,5,6,7,8,9,10,11}, while the controversies are discussed in \cite{12,13,14,15,16}.

Page and Wootters mechanism (PaW) is a proposal that considers time as a quantum degree of freedom \cite{17,18,19,20}. In this approach, all system variables are defined by quantum operators, and one of them is considered as the clock system that governs the dynamical evolution of the rest of the system. However, this proposal has faced criticisms from various perspectives, as discussed in \cite{7}. Specifically, the PaW conditional probability has been criticized for leading to incorrect localization probabilities for a relativistic particle, violating the Hamiltonian constraint, and resulting in incorrect transition probabilities. Fortunately, these objections have been addressed and resolved in subsequent works, as explained in \cite{33} and \cite{34}.

The objections raised against the PaW proposal \cite{7} can be restated as follows when applied to constrained systems. The main issue in the PaW approach lies in defining the variables as operators, particularly when it comes to defining a specific variable as a clock \cite{qt,MV}. In physical constrained systems, the observable variables are those that have Poisson brackets that vanish with the constraints. In other words, these variables remain invariant under the symmetries of the theory (gauge invariant). However, in systems where the Hamiltonian itself is one of the constraints, the other constraints that have vanishing Poisson brackets with the Hamiltonian do not evolve and are not suitable candidates for a clock. Page and Wootters attempted to address this problem by introducing kinematical variables that do not have vanishing Poisson brackets. As a result, these variables can evolve. However, this approach leads to a kinematical Hilbert space that cannot be interpreted probabilistically.

 A solution for the aforementioned problem is attempted in \cite{pu} through the utilization of a distinct set of physical quantities. These quantities are known as relational Dirac observables and can be likened to "evolving constants" \cite{5}. The proposal put forth in the aforementioned work applies the evolving constant approach to the construction of observables in canonical quantum gravity with PaW mechanism, utilizing a relational time for generally covariant systems.
 An alternative solution to address the criticisms of the PaW mechanism involves the formalization of measurements using the von Neumann prescription \cite{qt}, which is further extended to encompass generalized observable POVMs (positive-operator valued measures). By adopting this approach, all quantum predictions can be derived through the conditioning of the global, timeless state $|\Psi\rangle\rangle$. This approach offers the benefits of providing the correct quantum propagator and accurate statistics for measurements conducted at various points in time, advantages that are lacking in the PaW mechanism.
 
Nevertheless, despite the aforementioned alterations and criticisms towards the PaW mechanism, this approach presents itself as a sophisticated framework that facilitates an "evolution without evolution" \cite{pagewootters}. It boasts several promising attributes, including its compatibility with quantum gravity \cite{new1} and its practical nature, which holds great potential for experimental methodologies involving quantum clocks \cite{new2}.

In this article, our main focus is on the conceptual issues surrounding the PaW model. Unlike relational classical dynamics, the PaW model seems to have significant ambiguities. As highlighted in \cite{new3}, there is a notable "clock ambiguity" where multiple non-equivalent choices of the clock result in different Hamiltonians and drastically different dynamics for the rest of the universe. This implies that the timeless approach cannot be directly applied in the same manner as classical physics, as it fails to produce a unique Schrödinger equation for the rest of the universe.

However, it has been demonstrated that these clock ambiguities do not actually arise. To understand why, one must consider the essential properties that a subsystem must possess in order to function as a reliable clock, particularly its weak interaction with the rest of the system \cite{MV}.The PW model is further examined to determine the constraints that the state of the universe must adhere to in order for the model to be considered realistic. Additionally, it explores how the model incorporates a clear understanding of the flow of time. Consequently, the model becomes relevant to various unresolved issues and offers potential for new applications\cite{MV}.

 This paper investigates the time evolution of a system, the flow of time, the measurement problem, recording the history, and the consistent definition of the arrow of time in a curved spacetime near the horizon region of a black hole. The motivation for studying the PaW approach and these ambiguities in this region stems from the peculiar behavior of quantum effects, which can provide insights into the Page and Wootters timeless picture for describing the system's dynamics in curved spacetime. The generalization of the PaW approach  to curved spacetime has been frequently done in the last decade, making the study of ambiguities in this model highly beneficial\cite{nature,hamed,hadifarhad}. Additionally, the near horizon region of a black hole exhibits peculiar behaviors, further emphasizing the importance of applying the PaW approach to this region.
 
 However, the reader might question the rationale behind utilizing the PaW approach or any modified version of it in the vicinity of a black hole's near-horizon region. The region near the horizon is treated as if it were a flat space-time for an observer falling into the black hole, and as a Rindler space for a stationary observer. Consequently, it is logical to employ quantum field theory or quantum physics in this region, as it is a common practice in the literature of black hole physics. Given that the PaW approach falls within the framework of relational dynamics and addresses the issue of time, which is relevant in both classical physics and quantum mechanics \cite{endtime}, its application and investigation in the near-horizon region hold significant importance.
 
 Regarding this matter, we have examined the PaW method for the near-horizon area, which is a simple adaptation of the PaW mechanism in this region. This adaptation has already been carried out in Ref \cite{hadifarhad}. Furthermore, some of the criticisms of this mechanism have been reevaluated in this specific region, which has unique characteristics, such as the frozen vacuum near the horizon \cite{frozen}. This frozen vacuum is created when the non-local behavior of the black hole and its unitary evolution are taken into account, resulting in a violation of the equivalence principle of general relativity. In order to provide further clarification on this particular feature, it is important to highlight that the concept of complementarity \footnote{ The concept of complementarity asserts that the entirety of spacetime formed by a black hole source cannot be explored solely by a static or in-falling observer. However, when these two observers collaborate, they are able to investigate the entire spacetime.} in black holes appears to be contradicted by the presence of a firewall within the event horizon. This contradiction ultimately leads to a violation of the equivalence principle of General Relativity. To address this issue and maintain the principles of complementarity and equivalence, the $ER=EPR$ correspondence was proposed. (In this context, $ER$ refers to the Einstein Rosen bridge, while $EPR$ signifies the EPR paradox). Essentially, the firewall proposal aims to resolve the AMPS's paradox (abbreviated for Almheiri, Marolf, Polchinski, and Sully), but the $ER=EPR$ conjecture not only resolves the paradox but also upholds the equivalence principle and the complementarity of black holes. (For a more detailed explanation of the firewall, AMPS's paradox, and $ER=EPR$ conjecture, please refer below).

The AMPS claims that by accepting the complementarity of black hole the three following assumption leads to contradiction:
\begin{itemize}
        \item[1] An evaporating black hole preserves quantum information without destroying it. This property is called the unitary evolution of the black hole.
        \item[2] The in-falling observer crossing the event horizon of the black hole does not recognize any unusual property in horizon.  
        \item[3] Relativistic effective quantum field theory is  consistent for an static outside observer and she/he can work with the theory. 
        
        \end{itemize}
According to AMPS, the late radiation $B$ of an old black hole \footnote{An old black hole is the one which radiated half of its radiation away \cite{old}} is entangled with the early radiation $R_{B}$ in a maximal way. Assumptions 1 and 3 require that $B$ is entangled with a subsystem of $R_{B}$, while assumption 2 leads to entanglement between $B$ and the interior radiation $A$ of the black hole. These assumptions together violate the monogamy of entanglement \cite{mon1,mon2}, which states that if two quantum states are maximally entangled, then neither of them can be entangled with a third party. To resolve this paradox, AMPS proposes that the entanglement between $B$ and $A$ breaks down, releasing a high amount of energy in the black hole's horizon. This energy creates a firewall and extends the singularity to the horizon, indicating that there is no spacetime in the interior of the horizon  \cite{amps,amps2}.

The $ER=EPR$ conjecture presents itself as one of the potential solutions to AMPS's paradox. By proposing that the $ER$ bridge corresponds to an $EPR$ pair, this conjecture not only resolves AMPS's puzzle but also upholds the equivalence principle in the near-horizon region of a black hole. According to this idea, the $ER$ bridge is formed through the correlation of $EPR$ micro-states within two entangled black holes \cite{cool}. This conclusion is supported by the research conducted in \cite{cool1, cool2}. In simpler terms, a quantum system that exhibits $EPR$ correlation can be described by weakly coupled Einstein gravity, thereby explaining the nature of the $ER$ bridge. In fact, some physicists even propose that a quantum bridge of this kind exists for every singlet state, presenting a rather provocative notion. For a more comprehensive exploration of AMPS's paradox, along with various controversial ideas and potential solutions, please refer to \cite{antiamps}. However, when considering the $ER=EPR$ conjecture in the context of an older black hole, a violation of the equivalence principle arises within the "zone" \footnote{The term "zone" refers to an imaginary shell located at a distance approximately equal to the Schwarzschild radius from the black hole's horizon. In this paper, we use the terms "zone" and "near-horizon region" interchangeably.}. This violation occurs due to the acceptance of a specific and unconventional property within this region. The zone exhibits a peculiar behavior where the vacuum remains frozen, impervious to any attempts by an observer, whether in-falling or static, to excite it. This unique characteristic of the near-horizon region contradicts the principles of general relativity, thus challenging the equivalence principle of GR.

This article begins by reviewing the PaW approach in section II. In section III, the PaW approach is applied to the near-horizon region of a black hole. Subsequent sections highlight some of the controversial issues associated with this approach in the near-horizon region. While some of these problems have been resolved in the PaW mechanism in usual quantum mechanics in Hilbert space \cite{qt,MV}, there are still some issues, such as the time evolution of the system and flow of time, measurement problem, recording the history, and consistent definition of the arrow of time, that remain for the PaW approach in the black hole zone. Section IV specifically investigates the inability of the PaW mechanism to record history in the zone. The arrow of time is discussed in section V for the near-horizon region, while section VI examines the time ambiguity associated with the PaW approach in this region. Finally, the article concludes with a conclusion section.


\section{Review of PaW mechanism }
This section provides an overview of the PaW mechanism, which introduces the quantum operator for time in Hilbert space. The PaW approach aims to offer a coherent quantum representation of time through its conditional probabilities mechanism. The fundamental framework of this proposal is concisely examined in this section. The PaW approach consists of three distinct components: 
   
\begin{itemize}

\item[1]  The Hilbert space $\mathcal{H}_{R}$ of a system can be considered under an enlarged space as 
\begin{equation}
\mathcal{H}=\mathcal{H}_{C}\otimes \mathcal{H}_{R},      
\end{equation} 
where $\mathcal{H}_{C}$ is the space of ancillary system which is called the clock system. This system is equipped with canonical coordinates $T$ and $\Omega$ which resembles position and momentum and can be interpreted as time and energy indicator of evolving system. Canonical indicator of $T$ and $\Omega$ are defined as follows
\begin{equation}
        [T,\Omega]=i.
\end{equation}
The index $R$ in Hilbert space $\mathcal{H}_{R}$ stands for the "rest system" which we will use it when in PaW approach the system is divided into two subsystems of "clock" and "rest".

\item[2] One can describe the constraint operator of the system by following relation
\begin{equation}\label{gh}
\mathbb{H}=\hslash \Omega \otimes I_{R}+ I_{C} \otimes H_{R},    
\end{equation}
where $H_{R}$ is the system Hamiltonian and $I_{R}$ and $I_{C}$ are identity operators of $\mathcal{H}_{R}$ and $\mathcal{H}_{C}$, respectively. Also note that we present $H_{C}=\hslash \Omega$. 

\item[3] Next we define a physical vector of the system $|\Psi\rangle\rangle$ which has following feature
\begin{equation}\label{timless}
        \mathbb{H}|\Psi\rangle\rangle=0,
\end{equation}
double-Ket $|\Psi\rangle\rangle$ is defined on $\mathcal{H}_{C}\otimes \mathcal{H}_{R}$ .
The state $|\Psi\rangle\rangle$ is  physical vector of the model and is constrained by Hamiltonian $\mathbb{H}$ with null eigenvalue. $|\Psi\rangle\rangle$
is static state and does not evolve and represents the full history of the system $R$ consistently with Wheeler-Dewitt equation \cite{wde}.

 Note that, the state $|\Psi\rangle\rangle$ which is defined in Wheeler-DeWitt equation $(\ref{timless})$ is not only "Universe" state but also can be interpreted as a closed "universe" state with zero eigenvalue. The reason for it follows from the fact that by projecting $|\psi\rangle\rangle$ on the states $|\phi(t)\rangle_{C}=e^{-iH_{c}t/\hslash}|\phi(0)\rangle_{C}$ of the clock one obtains the states of the rest as $|\psi(t)\rangle_{R}:=_{C}\langle\phi(t)|\Psi \rangle\rangle =e^{-iH_{R}t/\hslash}|\psi(0)\rangle_{R}$ which indicates the proper evolution of subsystem $R$ by its own local Hamiltonian and its initial state is $|\psi(0)\rangle_{R}:=_{C}\langle\phi(0)|\Psi \rangle\rangle$. For an entangled state with an appropriate Hamiltonian the equation $(\ref{timless})$ can be written. For more information and experimental discussion about this refer to \cite{experimental}

One can derive the state $|\psi\rangle_{R}$ of the system by conditioning on solution $|\Psi\rangle\rangle$
with projection of generalized eigenvector time operator $T$ which is given by 

\begin{equation}
|\psi(t)\rangle_{R}=\ _{C}\langle t|\Psi\rangle\rangle,
\end{equation}
where $|\psi(t)\rangle$ is time dependent and the time projection of static state $|\Psi\rangle\rangle$. The time operator has following definition

\begin{eqnarray}
T|t\rangle_{C}=t|t\rangle_{C}, && _{C}\langle t^{'}|t\rangle_{C}=\delta(t-t^{'}). 
\end{eqnarray}
where $|t\rangle_{C}$ is the eigenvector of time operator $T$ with eigenvalue $t$ in clock system $C$. 
\end{itemize}
The crucial question that arises is whether it is possible to deduce the Schrödinger equation from the Wheeler-DeWitt equation (\ref{timless}). According to the PaW mechanism, one can derive the Schrödinger equation by converting equation (\ref{timless}) into position representation in $\mathcal{H}_{R}$ and following a series of steps. First $|t\rangle_{C}$ is multiplied from left over equation (\ref{timless}) 
\begin{equation}
        _{C}\langle t|\mathbb{H}|\Psi\rangle\rangle,
        \end{equation}
        then we have
        \begin{equation}
         i\frac{d}{dt} |\psi(t)\rangle=H_{R}|\psi(t)\rangle,
\end{equation}
where we have applied the following relation for $|\Psi\rangle\rangle$,
\begin{equation}
|\Psi\rangle\rangle=\int dt|t\rangle_{C}\otimes |\psi(t)\rangle_{R}.
\end{equation}
Although the wheeler-DeWitt equation (\ref{timless}) gives a timeless picture for universe system with eigenvectors $|\Psi\rangle\rangle$ by null eigenvalues, one can derive from it the Schr\"odinger equation which has time interpretation in its formalism.

A modification has been made to the PaW mechanism in order to accurately replicate the statistical outcomes of sequential measurements performed on a system at various points in time \cite{qt}. This revised approach provides a coherent depiction of time based on the PaW conditional probability mechanism and addresses the criticisms raised in the existing literature on the PaW picture.

\section{Frozen vacuum and quantized time in near-horizon region of a black hole}
In this segment, we examine the vicinity of the black hole's horizon and its frozen vacuum, followed by the application of the PaW mechanism to the near-horizon area. We revisit our earlier publication \cite{hadifarhad}, which is a simple adaptation of the PaW mechanism used in the previous section, specifically in the region adjacent to the horizon. It is worth noting that our methodology for the near-horizon region is a direct modification of \cite{qt}, with the same underlying principles.

 \subsection{Frozen vacuum}
 In this subsection, we will examine the frozen vacuum state of the near-horizon region of a black hole. Subsequently, we will utilize the modified PaW approach to analyze this particular region.

As stated in the introduction, the $ER=EPR$ correspondence plays a crucial role in resolving AMPS's paradox in black hole physics. This correspondence operates in the following manner: since the exterior states of the near-horizon region $B$ are maximally entangled with both the interior states $A$ and the early Hawking radiation $R_{B}$, it is necessary to break the entanglement between $B$ and either $A$ or $R_{B}$ in order to avoid violating the monogamy of entanglement. However, breaking the entanglement between $A$ and $B$ results in the formation of a firewall at the event horizon of the black hole, which violates the equivalence principle.

Nevertheless, the $ER=EPR$ correspondence asserts that the states $R_{B}$ are mapped to the states $A$. Consequently, based on this mapping, $B$ is entangled with only one state and is purified with $A$, while respecting the fact that $R_{B}$ corresponds to $A$. This resolves AMPS's paradox, and in reference \cite{frozen}, the mapping $A=R_{B}$ is referred to as the "donkey map". However, employing the donkey map leads to the emergence of a frozen vacuum in the near-horizon region.

Suppose that Alice, an observer falling into the horizon of an old black hole, has knowledge of the early radiation $R_{B}$. During her journey, she becomes aware of the inside states $A$ through the donkey map $(A=R_{B})$. By encountering the states $B$ in the near-horizon region, Alice can identify the purified states of $B$, which are essentially the same as $A$ according to $(A=R_{B})$ before entering the black hole. Consequently, Alice will encounter the vacuum state of the region close to the horizon as $|0\rangle_{ab}=\sum_{n=0}^{\infty}x^{n}|n\rangle_{a}|n\rangle_{b}$, where $|n\rangle_{a}$ and $|n\rangle_{b}$ represent the states inside and outside the horizon, respectively. The coefficient $x$ is defined as $x=e^{-\beta \omega /2}$
 		for modes with Killing frequency of the order of Hawking temperature and is
 		of order one.
 	
 	Suppose we consider a static observer named Bob, who is positioned near the horizon region outside the old black hole. Bob identifies the vacuum state as $|0\rangle_{ab}=\sum_{n=0}^{\infty}x^{n}|n\rangle_{a}|n\rangle_{b}$. Now, let's assume that Bob has the option to either excite region $B$ or solely observe a pointer. Initially, there is no interaction between the pointer and any of the subsystems involved.

 	\begin{equation}\label{pointer}
 		|\psi\rangle_{pR_{B}b}=|i\rangle_{p}\otimes\sum_{n=0}^{\infty}|n\rangle_{b}|n\rangle_{R_{B}}
 \end{equation}
 the pointer state is	$|i\rangle_{p}$ which has no interaction with any of the subsystems. The $ER=EPR$ correspondence with mapping $A=R_{B}$ yields
  \begin{equation}
 |0\rangle_{R_{B}}\rightarrow |0\rangle_{a},...
 |j\rangle_{R_{B}}\rightarrow |j\rangle_{a},...   
 \end{equation}
For simplicity we assume the old black hole  is billions of light years, then near-horizon region can be defined as several light years. In addition the curvature of near-horizon can be negligible and violation of semi-classical and equivalence principle may be expected.  

In the next step, the pointer $p$ measures the state $R_{B}$ so that the equation $(\ref{pointer})$ yields
\begin{equation}\label{pointer2}
	|\psi\rangle_{pR_{B}b}=\sum_{n=0}^{\infty}|n\rangle_{b}|n\rangle_{R_{B}}|n\rangle_{p}
\end{equation} 
A practical system cannot exist in isolation from its surroundings, and in this case, the role of the environment can be attributed to the pointer that measures the radiation states $R_{B}$. If we trace over the state $B$, the remaining $pR_{B}$ becomes a mixed state and loses its purity. Based on the aforementioned considerations, it is not possible for any mapping from $R_{B}$ to states $A$ to yield the vacuum state of the near-horizon region $|0\rangle_{ab}=\sum_{n=0}^{\infty}x^{n}|n\rangle_{a}|n\rangle_{b}$. Therefore, by including the environment $p$ for Hawking radiation $R_{B}$ the donkey map yields
\begin{equation}
|0\rangle_{R_{B}}|0\rangle_{p}\rightarrow |0\rangle_{a},...
|j\rangle_{R_{B}}|j\rangle_{p}\rightarrow |j\rangle_{a},...   
\end{equation} 
Then by suppressing the normalization factor the in-falling vacuum is
\begin{equation}\label{vacuumapp}
	|0\rangle_{ba} \propto \sum_{n=0}^{\infty}x^{n}|n\rangle_{b}|n\rangle_{a},
	\end{equation}
 Suppose that instead of measuring $R_{B}$, the pointer $p$ measures $b$, resulting in the same equation as equation $(\ref{pointer2})$. After this measurement, Bob, a stationary observer located one light year from the black hole horizon, disappears. Nine years later, Alice, a free-falling observer who is unaware of the previous measurement, experiences the vacuum near the horizon. Despite being close to the horizon, Alice does not observe anything special about the vacuum because she knows that $B$ can be purified by states $A$ that are identified by $R_{B}$. In other words, Alice was aware of $R_{B}$ before falling into the black hole, and therefore, she will not observe anything other than the in-falling vacuum $(\ref{vacuumapp})$ during her journey.

 However, if Alice becomes aware of Bob's knowledge, it results in a contradiction for her. This contradiction can be explained as follows: Let's assume that Bob meets Alice and shares the measurement $p$ of vacuum $B$ with her. If this meeting takes place, then Alice, while purifying $A(= R_{B})$ with B, faces a contradiction. The reason behind this is that in this particular situation, $B$ is not purified by Alice's view. In order to avoid this problem, Alice must not have the ability to stimulate the state in the near-horizon region on her own or recognize this stimulation by Bob. The only vacuum that she can encounter during her descent is the "frozen vacuum" $(\ref{vacuumapp})$.

 \subsection{PaW mechanism for near-horizon region}
In order to examine the operational definition of proper time \cite{a1,a2}, which is determined by a clock, it is necessary to have a physical system that functions as a clock. The additional Hilbert space $\mathcal{H}_{T}$ can be seen as an abstract purification space without any physical significance. However, in the modified PaW approach, $\mathcal{H}_{T}$ can be regarded as a physical system, and the clock can be idealized as being isomorphic to a particle on a line. The isomorphism between a clock and a particle on a line is discussed in \cite{11}. Nevertheless, attributing the quantum time degree of freedom as a dynamical degree of freedom connected to a physical system holds greater physical importance. The aforementioned definition of time can be measured by the time operator $T$, which is explained in section (2). In Newtonian mechanics, proper time can be interpreted as coordinate time (proper time = coordinate time), while in general relativity, coordinate time is equivalent to the proper time of a static inertial observer. For a more precise definition of proper and coordinate time, as well as the Hilbert space $\mathcal{H}_{T}$ in the near-horizon region using the modified PaW approach, please refer to \cite{hamed,nature}, which provides an average of $T$ to achieve a physical and classical interpretation of time.

However, in this section, we will consider the proper evolution of the inside of the black hole with a local Hamiltonian, according to the in-falling observer who falls into the horizon and recognizes a flat spacetime near the black hole's horizon. Now, we will review the PaW approach in the near-horizon region. The quantum state of the near horizon is the vacuum state as follows:
\begin{equation}\label{fv}
|\Psi\rangle\rangle=\sum_{n=0}^{\infty}x^{n}|n\rangle_{a}|n\rangle_{b},
\end{equation}
where $|n\rangle_{a}$ and $|n\rangle_{b}$ are the states of inside and outside of the horizon, respectively. As we explained above the coefficient $x$ is defined as $x=e^{-\beta \omega /2}$ for modes with Killing frequency of the order of Hawking temperature and is
of order one. 
In the literature of the black hole, $A$ and $B$ are used to indicate the inside and outside regions close to the horizon, respectively. We use this notation to indicate these regions. However, the clock system $C$ and the rest system $R$ are related to the PaW approach and are the notations for the clock and the rest.
The Hamiltonian of Hilbert space  $\mathcal{H}_{R}$   of the interior of  black hole horizon is written as follows
\begin{equation}\label{hrest}
H_{R}=-\sum_{n=0}^{\infty}x^{-n}|n\rangle_{aa}\langle n|.
\end{equation}
The ancillary system, which plays the role of the clock system, is the Hilbert space $\mathcal{H}_{C}$ of the exterior of the black hole's horizon. The Hamiltonian for this Hilbert space is:
\begin{equation}\label{hclock}
H_{C}=\sum_{n=0}^{\infty}x^{-n}|n\rangle_{bb}\langle n|.
\end{equation}
Notice that $H_{C}$ and $H_{R}$ indicate the local Hamiltonians of the exterior (clock system) and the interior (rest system) of the black hole's horizon, respectively. The Hamiltonian of the whole system is $\mathbb{H}=H_{C}\otimes I_{R}+ I_{C} \otimes H_{R}$, which has a Hilbert space representation of $\mathcal{H}=\mathcal{H}{C}\otimes\mathcal{H}{R}$. To obtain the Wheeler-DeWitt equation for this mode as in the previous section, the Hamiltonian $\mathbb{H}$ operates on the vacuum state $(\ref{fv})$ or, in other words, one can substitute equations (\ref{hclock}) and (\ref{hrest}) into equations (\ref{gh}) and (\ref{timless}), which leads to the following equation with a zero eigenvalue.

\begin{equation}
\left(\sum_{n=0}^{\infty}x^{-n}|n\rangle_{bb}\langle n| \otimes I_{R} - I_{C}\otimes \sum_{n=0}^{\infty}x^{-n}|n\rangle_{aa}\langle n|\right)|\Psi\rangle\rangle=0, 
\end{equation}

Where $|\Psi\rangle\rangle$ here is the frozen vacuum state $(\ref{fv})$, as we mentioned in the previous section, the vacuum state $|\Psi\rangle\rangle$ is interpreted as the universe state in the Wheeler-DeWitt equation. The reason for this choice is obvious from the explanations in the previous section and the following calculations.

 To derive the proper time evolution of the interior of black hole's states by PaW approach, the exterior observer can use 
her/his own subsystem states as follows
\begin{equation}
|\psi(t)\rangle_{b}=e^{-iH_{b}t/\hslash}|\psi(0)\rangle_{b}.
\end{equation} 
Then, to obtain the proper time evolution of the interior of black hole before falling into it, the observer can use the following 
equation
\begin{equation}\label{instate}
|\psi(t)\rangle_{a}=_{b}\langle \psi(t)|\Psi\rangle \rangle=e^{-iH_{a}t/\hslash}|\psi(0)\rangle_{a},
\end{equation} 
to obtain the time evolution of the inside of black hole. Note that here time is an emergent property. For more details and timeless picture of Wheeler-DeWitt equation refer to \cite{hadifarhad}.

\section{Recording history and flow of time in near-horizon region}

The PaW picture appears to lack the presence of time flow, yet the emergence of time flow becomes evident in this depiction through the division of the remaining universe system into three distinct sub-parts: the "observer" (interpreted solely as memory), the "observed," and a sequence of ancillas. These sub-parts have the capability to conduct measurements and retain their outcomes, thereby constructing a comprehensive history\cite{MV}.

In this section, we will examine the historical recording of the region near the horizon, which is crucial for understanding the passage of time in the PaW approach. It is important to note that the universe's state in the near-horizon region is a frozen vacuum state $(\ref{fv})$, denoted as $|\Psi\rangle\rangle \in \mathcal{H_{C}}\otimes\mathcal{H_{R}}$. Here, the states $A$ and $B$ belong to the subsystem Hilbert spaces of $\mathcal{H_{R}}$ and $\mathcal{H_{C}}$ respectively. It is worth mentioning that there is no distinction in defining $A$ or $B$ as the system of the clock. One can designate the interior state $A$ as the "clock" and the exterior state $B$ as the "rest of the universe".

 Now, suppose a static observer, Bob, hovers near the horizon region and performs a measurement on the "observed" to record what has happened. We assume that Bob's memory starts in a blank state $|r\rangle^{\otimes{N}}{M}$ at $t=0$. The "observed" is in the state $|\psi(t)\rangle$ and $|n\rangle{i}$ for $i = 1, 2, 3, ...$ are the eigenstates of the observable in subsystem $i$. Now, suppose that Bob and the "observed" evolve under an effective time-dependent Hamiltonian. Since Bob is in the exterior of the horizon $B$, we use the index $b$ instead of $i$. Bob measures the observable with an operator in his subsystem at time $t=1$. In this case, the local Hamiltonian operator $H_{R}$ (\ref{hrest}) would be suitable for the measurement. However, any operator can be used to do this work and the final result of this section does not change. This operation changes the state to
 \begin{equation}
        |n\rangle_{b}\otimes |r\rangle^{\otimes{N}}_{M} \rightarrow |n_{saw1}\rangle_{b} |r\rangle^{\otimes{N-1}}_{M} |1\rangle_{b},
 \end{equation}
 where $|n_{saw1}\rangle_{b}$ corresponds to state $|1\rangle_{b}$ which is the outcome of the measurement. A local permutation over ``observed'' can lead to, for example, the following state
  \begin{equation}
  |n_{saw1}\rangle_{b} |r\rangle^{\otimes{N-1}}_{M} |2\rangle_{b}.
  \end{equation}
  For the next step, Bob measures the state again which results in
  \begin{equation}
  |n_{saw1}\rangle_{b} |n_{saw2} \rangle_{b}|r\rangle^{\otimes{N-2}}_{M} |2\rangle_{b},   
  \end{equation} 
  where the state $|n_{saw2}\rangle_{b}$ corresponds to $|2\rangle_{b}$. Finally, a sequence of events is recorded by Bob such as eigenvalues of $|1\rangle_{b}, |2\rangle_{2}, |3\rangle_{3}, ...$ . He makes this record by a sequence of ancillas ($|n_{saw1}\rangle_{b}, |n_{saw2}\rangle_{b}, |n_{saw3}\rangle_{b}, ...$). These states can be interpreted as the states of the "pointers" in near-horizon region where Bob hovers there.

 Suppose Alice, as an observer falling into the black hole, is aware of the interior states $A$ through the $ER=EPR$ correspondence. This correspondence states that the interior states $A$ are a representation of the early Hawking radiation $R_{B}$, even without Alice meeting Bob and knowing the Hawking radiation states $R_{B}$ \cite{frozen, cool}.
  
   On the other hand, Alice is aware of the exterior states $B$ by her knowledge of the black hole, which claims that the exterior states $B$ are purified by the interior $A$. However, if Alice meets Bob, she will face a contradiction because, according to the donkey map ($A=R_{B}$), we have ($|n\rangle_{R_{B}} \rightarrow |n\rangle_{a}$). Alice, without meeting Bob, will confirm that the states $|n\rangle_{a}$ ($A$) purify the subsystem $B$, which are $|n\rangle_{b}$, and recognize the in-falling vacuum as $|\Psi\rangle\rangle=\sum_{n=0}^{\infty}x^{n}|n\rangle_{a}|n\rangle_{b}$, but with meeting Bob, she will admit that what purifies $A$ are the states ($|n_{saw1}\rangle_{b}|1\rangle_{b}, |n_{saw2}\rangle_{b}|2\rangle_{b}, |n_{saw3}\rangle_{b}|3\rangle_{b}, ...$). Anyway, this is a contradiction and is not possible because $|n_{saw1}\rangle_{b}|1\rangle_{b} \not=|n\rangle_{b}$. Therefore, Bob or even Alice herself are not able to record the history in the near-horizon region and what Alice recognizes is only the frozen vacuum $|\Psi\rangle\rangle=\sum_{n=0}^{\infty}x^{n}|n\rangle_{a}|n\rangle_{b}$.
  
  It should be noted that in this scenario, subsystem $B$ is regarded as the remainder of the system and is divided into three parts: "pointer" (ancillas), "observed," and "observer" (which solely comprises the memory system) that can be examined by Bob. Meanwhile, subsystem $A$ is viewed as the clock system that can be examined by Alice, an observer who is falling into the black hole.

  However, one can assign the role of the rest of the universe to subsystem $A$ and divide it, too. Anyway, recording the history for both subsystems is not possible because there is symmetry for subsystems $A$ and $B$. To explain it in more details, suppose the subsystem $A$ with the local Hamiltonian $H_{R}$, like before, subsystem $B$, is divided into $|n_{sawi}\rangle_{a}|i\rangle_{a}$ where $i=1, 2, 3,..$. Here, like before, the subsystem is partitioned into three sub-parts, but for simplicity we consider $A$ by dividing it into two sub-parts. The final result is independent of more partitioning. Like before, we know from the donkey map that the early Hawking radiation $R_{B}$ is a map of the interior space $A$, which can be represented by $|n\rangle_{R_{B}}\rightarrow |n_{sawi}\rangle_{a}|i\rangle_{a}$. Alice, according to her knowledge of the black hole, knows that $|n\rangle_{R_{B}}\rightarrow |n_{sawi}\rangle_{a} |i\rangle_{a}$, and also in her journey of falling into the black hole, by confronting states $B$, she knows which states purify $A$. In order to have a consistent interpretation of $A$, which on one hand comes from the early Hawking radiation $R_{B}$ by the map  $|n\rangle_{R_{B}}\rightarrow |n_{sawi}\rangle_{a} |i\rangle_{a}$ and on the other hand is the purified states of $B$, Alice must recognize the vacuum state of the near-horizon region as the frozen vacuum $(\ref{fv})$. In other words, whatever happens near the horizon region, the vacuum, is a frozen vacuum and nobody can create a particle there. For more information about this behavior, refer to \cite{frozen}.

   To summarize, while the PaW mechanism may have some applicability in certain regions, it is not feasible to record history in the near-horizon region. As a result, the PaW approach cannot fully address the criticisms outlined in \cite{MV}, and the concept of the flow of time remains contentious in this context. It should be noted that while the issue of recording history in accordance with the PaW approach in \cite{MV} has not been fully resolved in Hilbert space, it is clear that the method described in \cite{MV} is not suitable for the near-horizon region of a black hole, and this problem persists in that area.

  \section{No arrow of time in near-horizon region}
  In its framework, the PaW mechanism enables a time-reversal symmetric dynamical law similar to unitary quantum theory. Nonetheless, in order to discern the direction of time, it is necessary to introduce a quantity that steadily increases or decreases under a dynamical law. One proposal for the arrow of time in the PaW picture is the entanglement between the "observer" and the "observed" in the rest of the universe \cite{MV}. However, we demonstrate in this section that this definition does not hold true for the near-horizon region when employing the PaW approach.
  
  To establish the direction of time within the "zone," we proceed as follows. Let us consider the position of the pointer in the vicinity of the black hole.

We will define the progression of entanglement between the "pointer" and the "observed" as the arrow of time. While one could use the entanglement between the "observer" and the "observed" as a criterion for the arrow of time, we have chosen to focus on the pointer instead of the observer. This choice does not yield any disparity in the physical outcomes in this context.

Now, let us assume that the rest system $A$, which comprises the internal states of the black hole horizon, consists of both pointers and observed entities. The observed and the pointer in the premeasurement process are as follows.
  \begin{equation}\label{o1}
  |\psi\rangle_{pab} \approx  |0\rangle_{p}\otimes \sum_{n}^{\infty}x^{n}|n\rangle_{a}|n\rangle_{b},
  \end{equation}
    where $|0\rangle_{p}$ , $|n\rangle_{a}$ and $|n\rangle_{b}$ are states of pointer, interior and exterior of the horizon, respectively. The pointer $p$ is in premeasurement state and is not in any interaction with inside or outside of the horizon, yet.  In this case, the pointer can be written by inside states $A$ as follows
    \begin{equation}\label{o2}
        |\psi\rangle_{pab}=\sum_{n}^{\infty}x^{n}|n\rangle_{p}|n\rangle_{a}|n\rangle_{b}. 
    \end{equation}
    Equations (\ref{o1}) and (\ref{o2}) are called premeasurement.
The memory state of the observed $|n\rangle_{b}$ can also be represented by the pointer, serving as a record to keep track of events. In this context, the arrow of time can be explained by the increasing entanglement between the pointer and the observed, considering that entanglement never decreases in the rest of the system under the influence of $H_{R}$. However, it is not possible for the pointer to effectively keep track of the observed in the rest system, leading to a contradiction. This is because the in-falling observer encountering a black hole can only experience a frozen vacuum, as described by (\ref{fv}). In other words, the relationship between the pointer and the observed $|n\rangle_{p}|n\rangle_{a}$, in terms of the $ER=EPR$ correspondence, must be a representation of early Hawking radiation states $R_{B}$.
    \begin{equation}
    |n\rangle_{R_{B}} \rightarrow|n\rangle_{p}|n\rangle_{a}. 
    \end{equation}
    Conversely, Alice, an observer falling into the black hole, is aware of the maximal entanglement between $R_{B}$ and $B$ due to her knowledge of black holes. During her descent, Alice will encounter $B$ and knows that the inside state $A$ purifies it. However, if Alice detects the early Hawking radiation $R_{B}$ before the premeasurement of (\ref{o1}) and (\ref{o2}), she will face a contradiction as the expected inside state $A$ can only purify $B$ and cannot be $|n\rangle_{p}|n\rangle_{a}$. To avoid this contradiction, there must be no possibility of pointer $p$ recording the observed track. This implies that the entanglement criterion between pointer and observed is not applicable in this region. The frozen vacuum of the near-horizon region prevents the creation or excitation of any particle or pointer, making entanglement an unsuitable criterion for considering the arrow of time for PaW approach in the near-horizon region of the black hole.     
    
  \section{Near-horizon region and the unique clock} 
  
 In the PaW picture, another issue to be taken into account is the possibility of partitioning the entire Hilbert space of the universe into an infinite number of inequivalent tensor-product structures. This leads to multiple options for the clock-rest system, which is referred to as clock ambiguity \cite{MV}.
  
 First, we consider the clock ambiguity in detail and then indicate how it works in the near-horizon region. The clock ambiguity problem is as follows. One can choose a suitable orthonormal basis $|k\rangle$ in the overall Hilbert space $\mathcal{H}$,
  \begin{equation}
        |\psi\rangle=\sum_{k}\alpha_{k}|k\rangle.
  \end{equation} 
  where $|k\rangle$ can be written as $|t\rangle_{C}|\phi_{t}\rangle_{R}$ in a given tensor product structure $\mathcal{H} = \mathcal{H}_{C}\otimes\mathcal{H}_{R} $. Next, consider a different state of the universe, such as 
  \begin{equation}
        |\tilde{\psi}\rangle=\sum_{k}\beta_{k}|k\rangle.
  \end{equation} 
  We know that there is a unitary operator $S$ such that $|\tilde{\psi}\rangle=S|\psi\rangle$. Therefore we have
  \begin{equation}
        |\psi\rangle=\sum_{k}\beta_{k}S^{\dagger}|k\rangle=\sum_{k}\beta_{k}|\tilde{k}\rangle
  \end{equation}
  where $|\tilde{k}\rangle=S^{\dagger}|k\rangle.$
  
  The clock ambiguity claims that, as long as one can choose a basis $|\tilde{k}\rangle= |t\rangle_{\tilde{C}}|\phi_{t}\rangle_{\tilde{R}}$ with a different bipartite tensor product structure, considering the fact that there are countless such choices, this leads to a different description of the evolution of the rest. 
  
  There is no clock ambiguity because of following reasons:
  \begin{itemize}
        \item[Case one]: There is no interaction between subsystems $C$ and $R$: In this case, there is no ambiguity in clock definition.
        The tensor product structure of the system of clock and rest is $\mathcal{H_{C}}\otimes \mathcal{H_{R}}$ and since the clock and rest are non-interacting, the Hamiltonian is given by $H=H_{C}\otimes I_{R}+I_{C}\otimes H_{R}$. By applying the PaW approach the relative state $|\phi_{t}\rangle_{R}$ is evolved by a unitary evolution as follows
        \begin{equation}
                |\phi_{t}\rangle_{R}=exp(-iH_{R}t)|\phi_{0}\rangle_{R}.
        \end{equation}
        To make it clarify,  any state $|k\rangle \epsilon  \mathcal{H}$ subject
to a tensor product structure with a unitary map is written as follows
        \begin{equation}
                |k\rangle=\sum_{a,b}U^{k}_{a,b}|a\rangle_{C}|b\rangle_{R},
        \end{equation}
        where $|a\rangle_{C}|b\rangle_{R}$ are some basis states. Two tensor product structures are equivalent if and only if the elements 
        $U^{k}_{i,j}$ and $\tilde{U}^{k}_{a,b}$ are related by local unitary
operators $P$ and $Q$
        \begin{equation}\label{uu}
                U^{k}_{i,j}=\sum_{a,b}P^{a}_{i}Q^{b}_{j}\tilde{U}^{k}_{a,b}.
        \end{equation}
        These two tensor product structures are equivalent if $S = P \otimes Q$. Then, there is no any ambiguity in defining clock system.

        \item[Case one](in near-horizon region): This case  is applicable in the "zone" and there is no any ambiguity at all. To clarify it, assume the  interior and exterior states of the horizon as $A$ and $B$, respectively. When we restrict ourselves to $ER=EPR$
            paradigm for black hole, the frozen vacuum (\ref{fv}) is the only state in near-horizon region. One can define a local unitary
            operator for subsystems $A$ and $B$ separately. According to above explanations the ability of defining a local unitary operator for subsystems $A$ and $B$ leads to the claim that there is no any clock ambiguity \cite{MV}.

        \item[Case two]: There is interaction between subsystems $C$ and $R$: In this case, the new tensor product structure is not
        equivalent with old structure. The unitary operator $S$ has the form
        \begin{equation}
                S=exp\{-i(S_{C}+S_{R}+S_{CR})\},
        \end{equation}
        where $S_{CR}$ operates as an interaction term between two subsystems of the original tensor product structure. If the
        Hamiltonian of the system has following relation with $S$
        \begin{equation}
                [H,S]=0, ~~or~~ [H,S_{CR}]=0,
        \end{equation}
        then there is no any concern about ambiguity of the clock because $S$ would have a trivial local action on the state. When 
        $[H,S]\not=0$ then $H$ is the sum of two non-interacting terms of $C$ and $R$ in the tensor product structure which is defined by
        $U^{k}_{i,j}$. However, in new tensor product structure which is
defined by $\tilde{U}^{i}_{a,b}$ the Hamiltonian will have an 
        interacting term as follows
        \begin{equation}\label{interh}
                H=H_{\tilde{c}}\otimes I_{\tilde{R}}+I_{\tilde{C}}\otimes H_{\tilde{R}}+h_{\tilde{C}}\otimes h_{\tilde{R}}.
        \end{equation}
        Note that, the interaction term in equation $(\ref{interh})$ is fundamentally distinct from the interaction term discussed in section (IV) regarding the tripartite "clock-memory-rest" systems in the near-horizon region. In the latter case, the interaction occurs among the tripartite subsystems within $\mathcal{H_{R}}$, whereas in the former case, the interaction takes place between the clock and the rest of the universe systems ($\mathcal{H_{R}}$ and $\mathcal{H_{C}}$).
        
        Hence, due to the presence of interaction terms resulting in distinct tensor product structures, there is no room for confusion.Put differently, even though there are no local unitary operators satisfying condition (\ref{uu}), we still observe that $|\phi_{t}\rangle_{\tilde{R}} \neq exp(-iH_{\tilde{R}}t)|\phi_{0}\rangle_{\tilde{R}}$. Consequently, there is no ambiguity regarding the clock when there is interaction between the clock and the rest of the system. However, it is crucial to emphasize that we must establish a new clock and rest in order to operate within this new framework.

        \item[Case two] (in near-horizon region):When we apply the second case, that there are interaction terms between subsystems, in the "zone", we have to note that defining an interaction term by changing the basis and non-unitary tensor product structures is not possible. Therefore, without the possibility of non-unitary transformation in the new tensor product structure in the near horizon of the black hole, we would have a unique clock system. To clarify this issue, suppose $A$ and $B$ are the quantum states of inside and outside of the horizon, respectively. Like in the previous sections, $B$ is the clock system space and $A$ is the system of the rest. If there were an interaction between the clock ($B$) and the rest ($A$) systems, then the relative state of $B$ would change, for example, from $|n\rangle_{b}$ to $|n\rangle_{\tilde{b}}$. This change of basis for $B$ leads to a contradiction. Suppose an in-falling observer Alice, who knows the early Hawking radiation states $R_{B}$, wants to fall into the black hole. She is aware of $A$, the state of the inside, by a knowledge of the map ($A=R_{B}$) and also knowing that $B$ is purified by $A$. If there are interaction terms between $A$ and $B$, then Alice will face the states $|n\rangle_{\tilde{b}}$, which are not the purified states of $|n\rangle_{a}$ that she would expect to face.  
        
        To prevent any conflicting information, the sole state of the nearby horizon region is (\ref{fv}), and any alteration of the framework is limited to a unitary tensor product structure within this specific area. Consequently, there is no chance for any interaction between the clock and rest systems. Put differently, all tensor product structures hold the same significance within this zone. Hence, there exists only a single option for constructing a clock system, and this clock is  unique.

  \end{itemize}

  At the end of this section, we allude to the resolution of the issues discussed throughout this paper by employing a modified version of PaW mechanisms \cite{1,MV}. However, it should be noted that the criticisms directed towards the original form of the PaW formalism have not been entirely resolved. For a comprehensive understanding of the criticisms and potential solutions, references such as \cite{qt,MV,7,33,34,pu} can be consulted, where various modified forms of PaW have been developed. In this study, we have specifically focused on the modified forms proposed in \cite{qt,MV} for the near-horizon region. Furthermore, it is important to acknowledge the behavior of matter fields and gravitational fields in the vicinity of the black hole's event horizon. The presence of these fields necessitates the inclusion of interaction terms between the clock and the matter fields or gravitational field. In \cite{hamed}, the interaction between the clock and the center-of-mass of particles has been investigated within the backdrop of a black hole's near-horizon geometry, highlighting the influence of geometry on the PaW approach.

  \section{Conclusion}
  In this article, we have revisited the PaW mechanism in the vicinity of the black hole's horizon and examined certain properties and contentious concepts associated with this region. It has been suggested that capturing the system's history within the "zone" is unattainable. This system represents the vacuum state of the black hole's horizon. The act of recording history in the $ER=EPR$ paradigm for an observer falling into the black hole or a stationary observer in the near-horizon region gives rise to conflicting matters due to the frozen vacuum state.
  
 The second result is that the vacuum state of the near horizon does not exhibit an arrow of time as a result. Similar to the unitary quantum theory, the PaW mechanism also lacks an arrow of time. If we were to consider the arrow of time within this framework, it would need to be imposed by a separate postulate, where a specific property increases under the dynamical law. In the PaW approach, we divide one subsystem into two systems: the "observed" and the "pointer". The pointer is responsible for recording the evolution of the system, and through the entanglement property of these two systems, we have shown that it does not provide a consistent mechanism for recognizing the arrow of time in the near-horizon region. While entanglement is a useful feature for considering the arrow of time in the PaW approach, it is not applicable in this specific area of the black hole.
  
  Furthermore, the time ambiguity for the near horizon region was taken into consideration. It appears that the clock definition in the PaW mechanism does not have any ambiguity, unlike the common Hilbert space situation. The key finding of this section is that if there is no interaction term between the clock and the rest of the systems, then there is only one unique clock that can be used in the near horizon region of a black hole using the PaW approach. However, in the Hilbert space, even though there is no ambiguity in the clock system and its definition, a new clock system must be defined due to the interaction term. In the near-horizon region, there is only one unique clock because all changes in the basis are locally unitary, which ensures the unitarity of the entire system.
  
  \section*{Outlook}
  It is crucial to emphasize that the "zone"  is a highly unique region where the coexistence of quantum mechanics and general relativity is anticipated. Therefore, the exploration of quantum gravity in the near-horizon region proves to be more advantageous compared to investigating it in Hilbert space within curved space or under weak gravity conditions. The challenges posed by the time problem in quantum mechanics and canonical quantization of general relativity are subjects of great controversy. However, we believe that the near-horizon region offers a valuable setting for investigating these issues and potentially uncovering valuable insights into the nature of quantum gravity.

  \section*{Acknowledgments}
  This work is based upon research funded by Iran National Science Foundation (INSF) under project No 99033073.



\begin{thebibliography}{99}	
\bibitem{1}E.C.G. Stueckelberg, Remarks on the creation of pairs of particles in the theory of relativity, Helve. Phys. Acta 14, 322 (1941);E.C.G. Stueckelberg,La Mecanique du point materiel en theorie de relativite et en theorie des quanta, Helve. Phys. Acta 15, 23 (1942).
\bibitem{2}H.D. Zeh, Emergence of classical time from a universal wavefunction, Phys. Lett. A 116,9 (1986).     
\bibitem{3}R.P. Feynman, Simulating physics with computers,
Int. J. Theor. Phys. 21, 467 (1982); R.P. Feynman,
Quantum Mechanical Computers, Found. Phys. 16,
507 (1986); N. Margolus “Physics-like models of computation”, Physica D 10, 81 (1984).
\bibitem{4}C. Rovelli, Relational quantum mechanics, Int. J. of Theor. Phys. 35, 1637 (1996).
\bibitem{5}C. Rovelli, Time in quantum gravity: An hypothesis, Phys. Rev. D 43, 442 (1991).
\bibitem{pagewootters}D. Page and W. Wootters, Evolution without evolution,
Phys. Rev. D 27, 2885 (1983).
\bibitem{new1}C. J. Isham, in the NATO Advanced Study Institute Recent
Problems in Mathematical Physics, Salamanca, June 15,
1992, https://cds.cern.ch/record/241615/files/9210011.pdf.
\bibitem{new2} I. Ushijima, M. Takamoto, M. Das, T. Ohkubo, and H.
Katori, Cryogenic optical lattice clocks, Nat. Photonics 9,
185 (2015).
\bibitem{new3}A. Albrecht and A. Iglesias, The clock ambiguity and the
emergence of physical laws, Phys. Rev. D 77 063506
(2008).
\bibitem{MV}C. Marletto and V. Vedral, Evolution without evolution and without ambiguities, Phys. Rev. D 95, 043510 (2017).
\bibitem{6}D.N. Page and W.K. Wootters, Evolution without evolution: Dynamics described by stationary observables, Phys. Rev. D, 27, 2885 (1983).
\bibitem{nature}Alexander R. H. Smith, A. Ahmadi, Quantum clocks observe classical and
quantum time dilation, Nature Communications 11, 5360 (2020).
\bibitem{hamed}H. Hadi, K. Atazadeh, F. Darabi, Quantum time dilation in the
near-horizon region of a black hole, Physics Letters B 834 (2022) 137471.
\bibitem{hadifarhad} H. Hadi and F. Darabi, Time evolution of the inside of the black
hole’s horizon, accepted in EPJC.(2022)
\bibitem{7}K.V. Kuchar, Time and interpretations of quantum
gravity, Proc. 4th Canadian Conference on General Rel-
ativity and Relativistic Astrophysics, eds. G. Kunstatter,
D. Vincent, and J. Williams (World Scientific, Singapore,
1992), pg. 69-76.
\bibitem{8}J. Cotler, F. Wilczek, Entangled histories, Phys. Scr. 014004 (2016), arXiv:1502.02480 (2015).
\bibitem{9}Y. Aharonov, S. Popescu, J. Tollaksen, in Quantum Theory: A Two-Time Success Story (Springer, 2014) pg. 21-
36, arXiv:1305.1615 (2013).
\bibitem{10}D. Sels, M. Wouters, The thermodynamics of time, arXiv:1501.05567 (2015).
\bibitem{11}H. Salecker, E.P. Wigner, Quantum limitations of the measurement of space-time distances, Phys. Rev. 109, 571 (1958).
\bibitem{12}W.G. Unruh, R.M. Wald, Time and the interpretation of canonical quantum gravity, Phys. Rev. D 40, 2598 (1989).
\bibitem{13}E. Anderson, The problem of time in quantum gravity, arXiv:1009.2157v2 (2010).
\bibitem{14}K.V. Kuchar, in Quantum Gravity 2: a Second Oxford
Symposium ed. C.J. Isham, R. Penrose and D.W. Sciama
(Clarendon, Oxford 1981); K.V. Kuchar, in Conceptual
Problems of Quantum Gravity ed. A. Ashtekar and J.
Stachel (Birkhäuser, Boston 1991); C.J. Isham, in Inte-
grable Systems, Quantum Groups and Quantum Field
Theories ed. L.A. Ibort and M.A. Rodrıguez (Kluwer,
Dordrecht 1993), gr-qc/9210011; K.V. Kuchar, in The
Arguments of Time ed. J. Butterfield (Oxford University
Press, Oxford 1999).
\bibitem{15}R.D. Sorkin, Role of time in the sum-over-histories framework for gravity, Int. J. Theor. Phys. 33, 523 (1994).
\bibitem{16}D.N. Page, in Physical Origins of Time Asymmetry, eds.
J.J. Halliwell, et al., (Cambridge Univ. Press, 1993),
arXiv:gr-qc/9303020.
\bibitem{17}V. Vedral, Time, (inverse) temperature and cosmological inflation as entanglement, arXiv:1408.6965 (2014).
\bibitem{18}T. Banks, TCP, quantum gravity, the cosmological constant and all that..., Nucl. Phys. B 249, 332 (1985); R. Brout, On the concept of time and the origin of the cosmological temperature,
Found. Phys. 17, 603 (1987); R. Brout, G. Horwitz, D.
Weil, On the onset of time and temperature in cosmology, Phys. Lett. B 192, 318 (1987); R. Brout, Z. Phys.
B 68, 339 (1987).
\bibitem{19}R. Brunetti, K. Fredenhagen, M. Hoge, Time in quantum Physics: from an external parameter to an intrinsic observable, Found. Phys. 40,
1368 (2010); R. Brunetti, K. Fredenhagen, Time of occurrence observable in quantum mechanics, Phys. Rev. A
66, 044101 (2002).
\bibitem{20}In the context of cavity QED, a similar approach to quantizing a parameter was presented in: M. Wilczewski M.
Czachor Phys. Rev. A 80, 013802 (2009).
\bibitem{33} P. A. Höhn, A. R. H. Smith, and M. P. E. Lock, The trinity of relational
quantum dynamics. Preprint at https://arxiv.org/abs/1912.00033 (2019).
\bibitem{34} P.A.Höhn, A.R.H.Smith and M.P.E.Lock, Equivalence of Approaches to
Relational Quantum Dynamics in Relativistic Settings, Front. in Phys. 9,
181 (2021). doi:10.3389/fphy.2021.587083.
\bibitem{qt}V. Giovannetti, S. Lloyd, L. Maccone, Quantum time,         Phys. Rev. D 92, 045033 (2015), arXiv: 1504.04215
(2015)
\bibitem{a1}S.A. Basri, Operational foundation of Einstein's general theory of relativity, Rev. Mod. Phys. 37, 288 (1965).
\bibitem{a2} J.W. Kummer, S.A. Basri, Time in general relativity, Intern. J. Th. Phys. 2, 255 (1969).
\bibitem{von} J. von Neumann, Mathematical Foundations of Quantum
Mechanics (Princeton Univ. Press, 1955).
\bibitem{pu}R. Gambini, R.A. Porto, J. Pullin, S. Torterolo, Conditional probabilities with Dirac observables and the problem of time in quantum gravity
Phys. Rev. D 79, 041501(R) (2009).
\bibitem{frozen}R. Bousso,Violations of the equivalence principle by a non-locally reconstructed vacuum at the black hole horizon, Phys. Rev. Lett. 112, 041102
(2014). 
\bibitem{cool} J. Maldacena, L. Susskind, Cool horizon for entangled black holes,
Fortsch.Phys. 61 (2013) .
\bibitem{amps}A. Almheiri, D. Marolf, J. Polchinski and J. Sully, Black Holes: Complementarity or Firewalls? JHEP02(2013)062 arXiv:1207.3123 [hep-th].
\bibitem{old}D. N Page, Average entropy of a subsystem, Phys. Rev. Lett. 71,
1291 (1993), arXiv:gr- qc/9305007; D. N. Page, Black hole information,
arXiv:hep-th/9305040 (1993).
\bibitem{mon1}B. M. Terhal, Is entanglement monogamous?,IBM Journal of Research and
Development 48, (2004) arXiv:quant-ph/0307120 (2003).
\bibitem{mon2}M. Koashi and A. Winter, Monogamy of quantum entanglement and other
correlations, Phys. Rev. A 69, 022309 (2004), arXiv:quant-ph /0310037.
\bibitem{amps2}A. Almheiri, D. Marolf, J. Polchinski, D. Stanford, and J. Sully, An apologia
for firewalls, arXiv:1304.6483 (2013).
\bibitem{cool1}W. Israel, Thermofield dynamics of black holes, Phys. Lett. A 57, 107
(1976).
\bibitem{cool2}J. M. Maldacena, Eternal black holes in anti-de Sitter, JHEP 0304, 021
(2003) [hep-th/0106112].
\bibitem{antiamps}S. Lloyd, J. Preskill, Unitarity of black hole evaporation in the final-state
projection models, JHEP 08 126 (2014).
\bibitem{wde}B.S. DeWitt, Quantum theory of gravity. I. The canonical theory, Phys. Rev. 160, 1113 (1967).
\bibitem{experimental}Ekaterina Moreva, Giorgio Brida, Marco Gramegna, Vittorio Giovannetti,
Lorenzo Maccone, Marco Genovese, Time from quantum entanglement: an
experimental illustration, Phys. Rev. A 89, 052122 (2014).
\bibitem{endtime}J. Barbour, The End of Time: The Next Revolution in Physics
(Oxford University Press, New York, 1999).


\end{thebibliography}
\end{document}